\documentclass[prd,aps,showpacs,nofootinbib]{revtex4-1}

\usepackage{graphicx}
\usepackage{epsfig}
\usepackage[colorlinks=true, linkcolor=blue, citecolor=blue, urlcolor=blue]{hyperref}

\begin{document}

\title{Renormalization Group Invariance and Optimal QCD Renormalization Scale-Setting: A Key Issues Review}

\author{Xing-Gang Wu}
\email{wuxg@cqu.edu.cn}
\affiliation{Department of Physics, Chongqing University, Chongqing 401331, P.R. China}

\author{Yang Ma}
\email{mayangluon@cqu.edu.cn}
\affiliation{Department of Physics, Chongqing University, Chongqing 401331, P.R. China}

\author{Sheng-Quan Wang}
\email{sqwang@cqu.edu.cn}
\affiliation{Department of Physics, Chongqing University, Chongqing 401331, P.R. China}

\author{Hai-Bing Fu}
\email{fuhb@cqu.edu.cn}
\affiliation{Department of Physics, Chongqing University, Chongqing 401331, P.R. China}

\author{Hong-Hao Ma}
\email{mahonghao@cqu.edu.cn}
\affiliation{Department of Physics, Chongqing University, Chongqing 401331, P.R. China}

\author{Stanley J. Brodsky}
\email{sjbth@slac.stanford.edu}
\affiliation{SLAC National Accelerator Laboratory, Stanford University, Stanford, CA 94309, USA}

\author{Matin Mojaza}
\email{mojaza@cp3-origins.net}
\affiliation{CP3-Origins, Danish Institute for Advanced Studies, University of Southern Denmark, DK-5230}

\date{\today}

\begin{abstract}

A valid prediction for a physical observable from quantum field theory should be independent of the choice of renormalization scheme -- this is the primary requirement of renormalization group invariance (RGI). Satisfying scheme invariance is a challenging problem for perturbative QCD (pQCD), since a truncated perturbation series does not automatically satisfy the requirements of the renormalization group. In a previous review, which is published in Progress in Particle and Nuclear Physics ~\cite{Wu:2013ei}, we provided a general introduction to the various scale setting approaches suggested in the literature. As a step forward, in the present review, we present a discussion in depth of two well-established scale-setting methods based on RGI. One is the ``Principle of Maximum Conformality'' (PMC) in which the terms associated with the $\beta$-function are absorbed into the scale of the running coupling at each perturbative order; its predictions are scheme and scale independent at every finite order. The other approach is the ``Principle of Minimum Sensitivity" (PMS), which is based on local RGI; the PMS approach determines the optimal renormalization scale by requiring the slope of the approximant of an observable to vanish. In this paper, we present a detailed comparison of the PMC and PMS procedures by analyzing two physical observables $R_{e+e-}$ and $\Gamma(H\to b\bar{b})$ up to four-loop order in pQCD. At the four-loop level, the PMC and PMS predictions for both observables agree within small errors with those of conventional scale setting assuming a physically-motivated scale, and each prediction shows small scale dependences. However, the convergence of the pQCD series at high orders, behaves quite differently: The PMC displays the best pQCD convergence since it eliminates divergent renormalon terms; in contrast, the convergence of the PMS prediction is questionable, often even worse than the conventional prediction based on an arbitrary guess for the renormalization scale. PMC predictions also have the property that any residual dependence on the choice of initial scale is highly suppressed even for low-order predictions. Thus the PMC, based on the standard RGI, has a rigorous foundation; it eliminates an unnecessary systematic error for high precision pQCD predictions and can be widely applied to virtually all high-energy hadronic processes, including multi-scale problems.


\end{abstract}

\maketitle

\tableofcontents

\section{Introduction}
\label{sec:intro}

The setting of the renormalization scale of the QCD running coupling is one of the outstanding fundamental problems for perturbative QCD (pQCD) predictions; it is a key problem for obtaining high-precision predictions for high energy physics processes. In the pQCD framework, a physical quantity is expanded as a perturbative series in powers of the QCD running coupling. At any finite order, the renormalization scheme/scale dependence from the running coupling and the pQCD-calculable coefficient functions do not exactly cancel, leading to renormalization scheme/scale ambiguities. The elimination of such ambiguities is important for obtaining precise tests of the Standard Model (SM) at colliders such as the LHC and for increasing the sensitivity of experimental measurements to new physics~\cite{Wu:2013ei}.

It is essential to have an objective way of resolving the renormalization scale ambiguity. It has been conventional to choose a typical momentum transfer or a value which minimizes the contributions of the loop diagrams as the renormalization scale and to take an arbitrary range to estimate the uncertainties in the fixed-order QCD prediction. However, there is no guarantee that the actual pQCD prediction lies within the assumed range. In fact, the fixed-order prediction obtained by using a guessed scale depends heavily on the renormalization scheme which is itself arbitrary. It is often argued that by varying the scale, one can estimate the unknown contributions from higher-order terms. However, this procedure cannot expose the uncertainties from the non-$\beta $ terms in the perturbative series. Furthermore, the value of the effective number of quark flavors $n_f$ entering the QCD $\beta$-function is not determined by using conventional scale setting. Even worse, because of the presence of renormalon terms which diverge as ($n!\beta^{n}\alpha_s^n$)~\cite{Gross:1974jv,Lautrup:1977hs}, the convergence of a pQCD series based on a guessed scale becomes questionable for many processes. A review of the renormalon problem can be found in Ref.\cite{Beneke:1998ui}, in which it is shown that those renormalon terms can give sizable contributions to the theoretical estimates, such as $e^+e^-$ annihilation, $\tau$ decays, deep inelastic scattering, and hard scattering processes involving heavy quarks. Thus a careful treatment of the renormalon terms is also required for a reliable pQCD prediction. In the literature, the large $\beta_0$ approximation~\cite{Neubert:1994vb,LovettTurner:1994hx,LovettTurner:1995ti,Ball:1995ni} has been suggested to deal with the renormalon terms. For example, as a first step toward studying the renormalons in the non-relativistic QCD factorization formalism, the large $\beta_0$ approximation has been applied to deal with the electromagnetic annihilation decays of the quarkonium~\cite{Braaten:1998au}.

The running behavior of the coupling constant is governed by renormalization group equations (RGEs)~\cite{Bogolyubov:1980nc}, and valid predictions for physical observables must satisfy renormalization group invariance (RGI)~\cite{Petermann:1953wpa,GellMann:1954fq,Peterman:1978tb,peter2,Callan:1970yg,Symanzik:1970rt}; i.e., the prediction for a physical observable must be independent of the choice of renormalization scheme. This is the key requirement of the renormalization group.

Thus, a primary problem for pQCD is how to set the renormalization scale so as to obtain the most accurate fixed-order estimate while satisfying the principles of the renormalization group. Two approaches based on RGI have been suggested since the 1980's. One is the Brodsky-Lepage-Mackenzie (BLM) method~\cite{BLM}, which has been further developed as the ``Principle of Maximum Conformality (PMC)~\cite{pmc1,pmc2,pmc3,pmc5,BMW,BMW2}. The other approach is the ``Principle of Minimum Sensitivity (PMS)"~\cite{PMS1,PMS2,PMS3,PMS4}.

Since the running behavior of the coupling constant is governed by the $\beta$-function of RGE, the $\beta$-terms of a process can be used to determine the optimized ``physical" scales of the process. This procedure stimulated the suggestion of BLM/PMC. In the BLM/PMC method, all terms associated with the $\beta$-function are absorbed into the scale of the running coupling at each perturbative order via a step-by-step way, leaving a series with coefficients identical to that of the corresponding conformal theory with $\beta=0$; the resulting predictions are then scheme and scale independent at every finite order. Since the invention of BLM, it has been widely accepted in the literature for dealing with high energy processes. Moreover, it is found that by applying the BLM for determining the effective scale, the predictive power of lattice perturbation theory can be greatly enhanced~\cite{Lepage:1992xa}.

In the original BLM paper, it was proposed that one can use the occurrence of $n_f$-terms in the series as a guide to identifying the $\beta$ terms. This procedure is easily implemented at low orders; however, at high orders, the $n_f$ terms can also arise from loops which are ultraviolet finite and are not associated with the $\beta$ function. Thus the key problem for extending the BLM method to high orders is how to set the $n_f$ and $\beta$ correspondence correctly. The PMC is designed for this purpose at any order. The PMC provides the underlying principle and rigorous foundation for BLM, giving a systematic method for unambiguously distinguishing the $\beta$ versus non-$\beta$ ``conformal terms". The PMC thus respects RGI; the final expression is naturally scheme and scale independent at any finite order since all non-conformal $\beta$-terms are absorbed into the coupling constant. The PMC fixes the scales correctly and individually at each perturbative order. The resulting renormalization scales depend on the choice of the renormalization scheme. For example, the renormalization scales in different $R_\delta$ schemes, $R_{\delta_1}$ and $R_{\delta_2}$, differ only by a factor $e^{(\delta_1-\delta_2)/2}$~\cite{BMW2}. The PMC scale relations also ensure the scheme independence of the pQCD predictions among different schemes. The renormalon problem is also avoided by the PMC, and the pQCD convergence is thus greatly improved.

A different way for applying BLM scale setting to higher orders has also been suggested, i.e. the ``seBLM approach"~\cite{Mikhailov:2004iq,Kataev:2010du}. However, the main purpose of seBLM is to improve the pQCD convergence, in which the large $\beta_0$ approximation has been adopted as a guide to deal with the pQCD series. Comparisons of PMC with a modified seBLM version up to four-loop QCD corrections can be found in Refs.\cite{Wang:2013bla,Ma:2015dxa}.

The PMS determines the optimal renormalization scale by requiring the slope of the approximant of an observable to vanish. In effect, the PMS breaks the standard RGI but introduces instead a local RGI. The local RGI requires the fixed-order series to satisfy the RGI at the renormalization point. Thus, in distinction to the PMC, the resulting PMS scale and scheme are both fixed in order to achieve the most stable pQCD prediction over the choices of renormalization scales and schemes. Since it breaks the standard RGI, the PMS does not satisfy the self-consistency conditions of the renormalization group, such as reflectivity, symmetry and transitivity, as discussed in Ref.\cite{pmccolloquium}. In some cases, the predicted PMS scale does not have the correct physical behavior. For example, for the jet production via $e^+ e^-$ annihilation, the predicted PMS scale rises anomalously without bound with decreasing small jet energy~\cite{Kramer1,Kramer2} and thus small gluon virtuality. The PMS, however, provides an intuitive way to set the renormalization scale, and its predictions tend to be steady over the changes of renormalization scheme/scale around the determined renormalization point. The PMS applies the local RGI step-by-step to set the PMS scale, and the resulting RGI coefficients at each perturbative order are based on its own self-consistency conditions~\cite{PMS2}.

Many attempts have been tried to solve the renormalization scale and renormalization scheme ambiguities. In addition to the above mentioned PMC and PMS, another method, the renormalization-group-improved effective coupling method (or the so-called Fastest Apparent Convergence (FAC))~\cite{Grunberg:1980ja,FAC3,FAC4}, or the closely related Complete Renormalization Group Improvement (CORGI) approach~\cite{Max1,Max2,Max3}, has also been suggested. The main purpose of the FAC is to improve the pQCD perturbative series by requiring all higher-order terms beyond leading order to vanish. However, this method in effect redefines the renormalization scheme as an effective charge for each observable, i.e. all the known-type of higher-order corrections are designed to be absorbed into an effective coupling through RGE in order to provide a reliable estimation. A detailed introduction to various scale setting methods can be found in a recent review~\cite{Wu:2013ei}. In this paper, we shall concentrate on the renormalization scale setting solutions based on the RGI principle. For illustration, we shall present a detailed comparison of PMC and PMS predictions for two high-energy processes up to four-loop level; specifically the processes $e^+ e^- \to {\rm hadrons}$ and the Higgs decay $H\to b\bar{b}$. This comparison illuminates the merits and differences of the two RGI methods, PMC and PMS, for confronting the scale-setting problem.

The remaining parts of the paper are organized as follows: In Sec.\ref{pQCDexpansion}, we present a general argument for the form of pQCD expansions and the analysis of the scale-setting problem. In Sec.\ref{comparing}, we present a comparison of PMC and PMS scale settings. In Sec.\ref{sec:pmc}, we present the standard RGI and the formulae for PMC up to four-loop level. In Sec.\ref{sec:pms}, we show how one can implement local RGI and obtain the formulae for PMS up to four-loop level. In Sec.\ref{sec:comp}, we present our numerical results for the two processes $e^+ e^- \to {\rm hadrons}$ and $H\to b\bar{b}$. A detailed comparison of the PMC and PMS predictions for the annihilation ratio $R_{e^+e^-}$ and the Higgs decay width $\Gamma(H\rightarrow b\bar{b})$ up to four-loop level, together with the predictions using the conventional scale setting, is then presented. A method for estimating the ``unknown" higher order pQCD corrections is also been presented. Sec.\ref{sec:summary} is reserved for the  summary.

\section{Expansions in Perturbative QCD and the Renormalization Scale Setting Problem} \label{pQCDexpansion}

Because of the asymptotic freedom property of quantum chromodynamics (QCD)~\cite{qcd1,qcd2}, a high-energy physical observable ($\varrho$) can be expanded in perturbative series in powers of the strong running coupling $\alpha_s(\mu)$. For simplicity we shall consider the series when the quark masses vanish. At the $n$-th order, we have
\begin{equation} \label{phyvalue}
\varrho_n = {\cal C}_0 \alpha_s^p(\mu) + \sum_{i=1}^{n}{\cal C}_i(\mu) \alpha_s^{p+i}(\mu),
\end{equation}
where $\mu$ stands for the renormalization scale of the running coupling $\alpha_s$, ${\cal C}_0$ is the tree-level term and $p$ is the power of the coupling associated with the tree-level term, ${\cal C}_1$ the one-loop correction; etc. Typically, the higher-order coefficients ${\cal C}_{i\geq1}$ are ultraviolet divergent which must be regulated and removed by a renormalization procedure. The terms which are associated with the renormalization of the running coupling involve contributions to the $\beta$ function, the logarithmic derivative of $\alpha_s$. The remaining terms are identical to a ``conformal" theory with $\beta =0$. Because of RGI, a physical prediction, calculated up to all orders, should be independent of the choice of renormalization scheme and scale. However, at any finite order, the renormalization scheme/scale dependence from $\alpha_s(\mu)$ and ${\cal C}_i(\mu)$ usually do not exactly cancel, leading to the well-known renormalization scheme/scale ambiguities, cf.Refs.\cite{Grunberg:1980ja,FAC3,PMS1,PMS2,PMS3,PMS4,BLM} and references therein. Such ambiguity usually constitutes a systematic error for a pQCD prediction.

In the case of conventional scale setting, one simply guesses the renormalization scale and varies it over an arbitrary range. It is a common belief that the effects of the renormalization scale uncertainty will be reduced as one proceeds to higher-and-higher order calculations. However, this {\it ad hoc} assignment of renormalization scale and its range introduces an important and persistent systematic and scheme-dependent error in the theoretical predictions. It should be emphasized that the variation of the renormalization scale can only provide a rough estimate of the higher-order non-conformal terms but not the conformal ones. Uncanceled large logarithms as well as the factorial growth of the ``renormalon" terms in higher orders will provide sizable contributions to the theoretical predictions and largely dilute the perturbative nature of the expansion series. As an example, the large next-to-leading order (NLO) contributions observed in the literature for the heavy quarkonium productions/decays are mainly caused by such renormalon terms. It is sometimes argued that the correct renormalization scale for the fixed-order prediction can be decided by comparing with the experimental data; however, this procedure is process-dependent, and it greatly depresses the predictive power of pQCD.

One may expect that the uncertainties introduced from the conventional renormalization scheme/scale dependence can be eliminated if one can find the optimal behavior of the coupling constant via a systematic and process-independent way. As mentioned in the Introduction, various scale setting procedures have been proposed in the literature. The PMC and PMS methods are designed to eliminate the scheme/scale ambiguity and to find the optimal behavior of the coupling constant; however, they have quite different consequences due to different starting points, and they may or may not achieve the desired goals. In the following sections, we shall concentrate our attention on the PMC and PMS methods in which the basic RGI principle has been adopted with the hope to eliminate the renormalization scheme/scale dependence fundamentally and simultaneously.

\section{Comparing the Principles of Maximum Conformality and Minimal Sensitivity}
\label{comparing}

The scale dependence of the strong coupling constant is controlled by its RGE. The PMC provides the underlying principle for BLM; it respects the standard RGI and improves the perturbative series by absorbing all $\beta$-terms governed by RGE into the coupling constant. This procedure is identical to the Gell-Mann-Low procedure in QED whereby all proper and improper vacuum polarization contributions are absorbed into the photon propagator by choosing the scale of $\alpha(q^2)$ as photon virtuality~\cite{GellMann:1954fq}. Early criticisms on the effectiveness of NLO BLM prediction to eliminate scheme dependence~\cite{Celmaster:1982zj} have been softened by using the method of effective charges~\cite{Grunberg:1984py,Grunberg:1992mp}, and later on have been clarified by the development of commensurate scale relations (CSRs) suggested in Ref.~\cite{Brodsky:1994eh}; these relations, which relate physical observables to each other, ensure that the NLO BLM/PMC predictions are independent of the choice of the renormalization scheme for any observable. Reactions with multi-gluon couplings are more difficult to analyze using BLM because quark loops appear in high-order corrections to the multi-gluon vertex as well as in the propagator insertions~\cite{Binger:2006sj}. Scale setting for the BFKL Pomeron intercept provides such an example~\cite{pomeron1,Zheng:2013uja,Caporale:2015uva}.

The PMC provides the general procedure which underlies those CSRs. A rigorous demonstration of the scheme independence at any fixed-order can be obtained by using the $R_\delta$-scheme~\cite{BMW,BMW2}, a systematic generalization of the minimal subtraction renormalization scheme. The PMC procedure is identical to the Gell-Mann-Low procedure in the limit $N_C \to 0$ at fixed $\alpha=C_F \alpha_s$ with $C_F=(N_c^2-1)/2N_c$~\cite{qedlimit1,qedlimit2}. Since the pQCD series is identical to the series of a conformal theory with $\beta=0$, the PMC prediction has the remarkable feature that it is scheme independent at every finite order. The PMC satisfies all the self-consistency conditions of the renormalization group, such as reflectivity, symmetry and transitivity~\cite{pmccolloquium}. Since the running coupling sums all of the $\beta$ terms, the divergent ``renormalon" series does not appear in the PMC prediction, allowing the convergence of the pQCD series.

The PMS is designed to solve the renormalization scheme and scale ambiguity by applying the so-called ``local RGI"; one requires the fixed-order series to satisfy the RGI at the renormalization point. Since it breaks the standard RGI, the PMS does not satisfy the self-consistency conditions of the renormalization group, such as reflectivity, symmetry and transitivity, as discussed in Ref.\cite{pmccolloquium}. It, however, provides an intuitive way to set the renormalization scale, and its predictions tend to be steady over the changes of scheme/scale around the determined renormalization point. The PMS applies the local RGI step-by-step to set the PMS scale, and the resulting RGI coefficients at each perturbative order are based on its own self-consistency conditions~\cite{PMS2}. For example, at $n$-th order, we have
\begin{equation}\label{pmsconst}
\partial \varrho_{i} /\partial {\rm (RS)}={\cal O}(\alpha^{i+1}_s)
\end{equation}
where $i=(1,2,\cdots,n)$ and ${\rm RS}$ stands for either the scale or the scheme parameters. Recently, the PMS has been extended up to four-loop order~\cite{PMS5,Ma:2014oba}, the key point of which is to fix the local RG invariants at each order.

Both the PMC (and its precursor BLM) and PMS are well-known and have been applied to many high-energy processes. Most of the previous analysis of PMC/BLM and PMS have only dealt with predictions at the one-loop level. However, in recent years, due to the significant development of new loop calculation technologies, many interesting high-energy processes have been calculated up to two-loop, three-loop, or even up to four-loop level. Thus, we are facing the opportunities for testing PMC and PMS at a much higher confidence level. We emphasize that the PMC and PMS are based on different theoretical principles; e.g., the standard RGI versus local RGI, respectively; thus, the predictions of PMC and PMS behave quite differently. A comparison of PMC and PMS, together with conventional scale setting up to high-loops, is important, and this is one of the main purposes of this paper.

\section{PMC and Standard RGI} \label{sec:pmc}

As has been pointed out in Refs.\cite{PMS1,PMS2,pmccolloquium,Lu:1992nt}, it is convenient to introduce extended RGEs for determining the running behavior of the coupling constant. For this purpose, one can define a universal coupling constant $a(\tau_{\cal R}, \{c^{\cal R}_i\})$ which satisfies the following extended RGEs,
\begin{equation}
\beta(a,\{c^{\cal R}_i\}) =\frac{\partial a}{\partial \tau_{\cal R}} = -a^2 \left[1+ a +c^{\cal R}_2 a^2+c^{\cal R}_3 a^3 +\cdots \right] \label{scale}
\end{equation}
and
\begin{equation}
\beta_n(a,\{c^{\cal R}_i\}) = \frac{\partial a}{\partial c^{\cal R}_n} = -\beta(a,\{c^{\cal R}_i\}) \int_0^{a} \frac{ x^{n+2} dx}{\beta^2(x,\{c^{\cal R}_i\})} \, , \label{scheme}
\end{equation}
where for any given ${\cal R}$-renormalization scheme, the coefficients are $c^{\cal R}_i = {\beta^{\cal R}_i \beta_0^{i-1}} / {\beta^i_1}$ $(i=2, 3, \cdots)$. We have implicitly used the scheme-independent $\beta_0$ and $\beta_1$ to rescale the coupling constant and the scale-parameters, i.e. $a(\tau_{\cal R},\{c^{\cal R}_i\}) = \frac{\beta_1} {4\pi\beta_0}\alpha^{\cal R}_s(\tau_{\cal R},\{c^{\cal R}_i\})$ and $\tau_{\cal R} =\frac{\beta^2_0}{\beta_1} \ln\mu^2|_{\cal R}$. The scale-equation (\ref{scale}) determines the running behavior of the universal coupling function, whose solution can be derived in a recursive way~\cite{pmc2}. The scheme-equation (\ref{scheme}) determines the relation of the coupling functions among different schemes, whose solution can be achieved via a perturbative expansion in the QCD coupling.

The RGI principle requires that the prediction for a physical observable should be independent of the choice of the renormalization scheme or initial scale~\cite{Petermann:1953wpa,GellMann:1954fq,Peterman:1978tb,peter2,Callan:1970yg,Symanzik:1970rt}. As suggested in Refs.\cite{Grunberg:1980ja,FAC3}, if an effective coupling $a(\tau_{\cal R},\{c^{\cal R}_i\})$ corresponds to a physical observable, then it should be independent of any other scale $\tau_{\cal S}$ and any scheme parameters $\{c^{\cal S}_j\}$,
\begin{eqnarray}
\frac{\partial a(\tau_R,\{c^R_i\})}{\partial \tau_S} &\equiv& 0 \, , \label{inv-scale} \\
\frac{\partial a(\tau_R,\{c^R_i\})}{\partial c^{S}_j} &\equiv& 0 \, .\label{inv-sch}
\end{eqnarray}
Based on the RGEs (\ref{scale},\ref{scheme}), we can obtain a direct deduction of Eqs.(\ref{inv-scale},\ref{inv-sch}) for an $n$-th order estimate~\cite{pmccolloquium},
\begin{widetext}
\begin{eqnarray}
\frac{\partial a(\tau_R,\{c^R_i\})}{\partial \tau_S} =\frac{\partial^{(n+1)} a(\tau_S,\{c^S_i \})} {\partial\tau_S^{(n+1)}}\frac{\bar{\tau}^{n}}{n!} + \sum_{i} \frac{\partial^{(n+1)}a(\tau_S,\{c^S_i \})} {{\partial c^S_i}\partial\tau_S^{(n)}}\frac{\bar{\tau}^{n-1}\bar{c}_{i}}{(n-1)!} + \cdots , \label{rgi-use}
\end{eqnarray}
\end{widetext}
where ${\cal R}$ and ${\cal S}$ stands for two renormalization schemes, $\bar{\tau}=\tau_{\cal R} -\tau_{\cal S}$ and $\bar{c}_i =c^{\cal R}_i - c^{\cal S}_i$. If setting $n\to\infty$, the theoretical estimate for the physical observable $a(\tau_R,\{c^R_i\})$ will be independent of any other scale $\tau_S$. Similarly, by taking the first derivative of $a(\tau_R,\{c^R_i\})$ with respect to $c^{S}_j$, one can also obtain the scheme-invariance equation (\ref{inv-sch}) for $n\to\infty$. Thus, the RGI Eqs.(\ref{inv-scale},\ref{inv-sch}) tell us that, I) if we could sum all types of $c^{\cal S}_i$-terms (or equivalently the $\{\beta^{\cal S}_i\}$-terms) into the coupling constant, then the final prediction of $a(\tau_R,\{c^R_i\})$ will be independent of any choice of scheme and scale; II) There can be residual scale dependence for a fixed-order estimate; e.g., if $n\neq\infty$, the right-hand of Eq.(\ref{rgi-use}) is non-zero.

Note that by setting $\bar{c}_i \equiv 0$ ($i=1,2,\cdots$), we can obtain a scale-expansion series for the coupling constant expanding over itself but specified at another scale; i.e.,
\begin{widetext}
\begin{eqnarray}
a(\tau_{\cal R},\{c^{\cal R}_i\})&=& a(\tau_{\cal S},\{c^{\cal R}_i \})+ \left(\frac{\partial a(\tau_{\cal S},\{c^{\cal R}_i \})}{\partial \tau_{\cal S}}\right) \bar{\tau} + \frac{1}{2!}\left(\frac{\partial^2 a(\tau_{\cal S},\{c^{\cal R}_i \})}{\partial \tau_{\cal S}^2}\right) \bar{\tau}^2 +\frac{1}{3!}\left(\frac{\partial^3 a(\tau_{\cal S},\{c^{\cal R}_i \})}{\partial \tau_{\cal S}^3}\right) \bar{\tau}^3 +\cdots . \label{betaseries}
\end{eqnarray}
\end{widetext}
Using the scale-equation (\ref{scale}), the above equation can be rewritten as a perturbative series of $a(\tau_{\cal S},\{c^{\cal R}_i \})$, whose coefficient at each order is a $\{\beta^{\cal R}_i\}$-series.

In summary, the standard RGI indicates that if one can resum all the known-type of $\beta$-terms into the coupling constant, and at the same time suppress the contributions from those unknown $\beta$-terms, then one may solve the conventional scheme and scale ambiguity. This observation is the underlying motivation of PMC.

\begin{figure}[tb]
\includegraphics[width=0.6\textwidth]{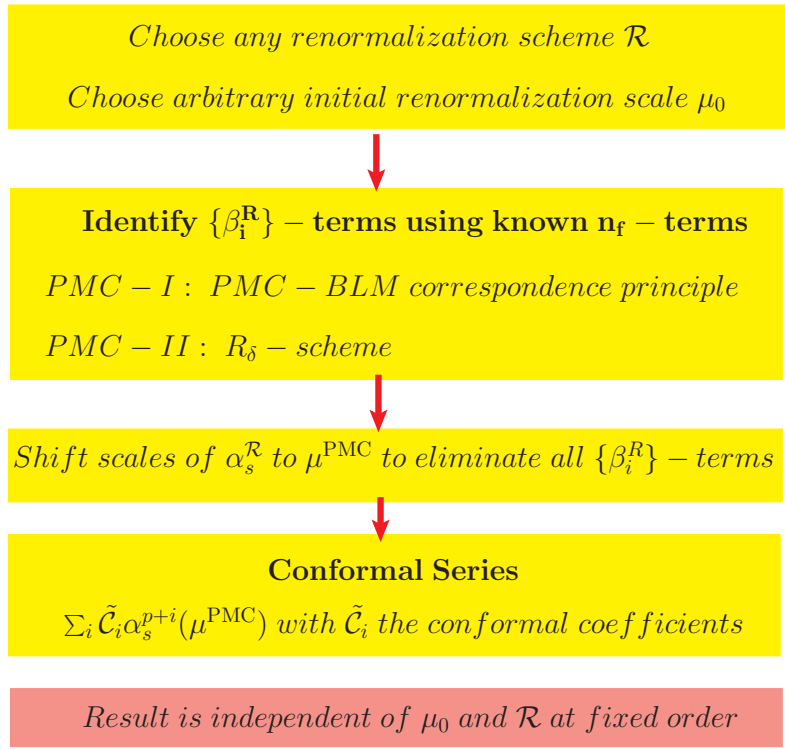}
\caption{A ``flow chart" which illustrates the PMC procedure. Two ways, named as PMC-I and PMC-II, are suggested to absorb the $\beta$-terms into the coupling constant and the final resultant is conformal and independent of the initial choice of scheme and scale. }
\label{fig1}
\end{figure}

The PMC provides an unambiguous and systematic way to set the optimized renormalization scale at each finite orders. A ``flow chart" which illustrates the PMC procedure is presented in FIG. \ref{fig1}. We first perform a pQCD calculation by using any renormalization scheme (usually $\overline{\rm MS}$-scheme) at an arbitrary initial scale (its value should ensure the perturbative calculation applicable). Then, we arrange all the coefficients at each perturbative order into $\beta$-terms or non-$\beta$-terms depending on whether they are pertained to the renormalization of the coupling constant. The $\beta$-terms are then absorbed into the coupling constant in an order-by-order manner. New $\beta$-terms will occur at each perturbative order, so the PMC scale at each order is in general distinct.

In practice all the $\beta$-terms involve UV-divergent light-quark loops, e.g. the $n_f$-terms; thus how to correctly relate the $\beta$-terms to the $n_f$-terms is the key problem of PMC. Two ways have been suggested to derive a one-to-one correspondence between the $\beta$-terms and the $n_f$-terms, one is based on the PMC-BLM correspondence principle (we call it as PMC-I)~\cite{pmc2,pmc5} and the other one is based on the so-called $R_\delta$-scheme (we call it as PMC-II)~\cite{BMW,BMW2}. The PMC-I and PMC-II methods can be conveniently extended up to any perturbative order. In the following, we present the main ideas and provide the formulae up to four-loop level.

\subsection{PMC-I: Achieving the goal of PMC via PMC-BLM correspondence principle}

The PMC-I approach uses the PMC-BLM correspondence principle~\cite{pmc2} to obtain an unambiguous relations among the $\beta$-terms and the relevant $n_f$-terms at each order of perturbative theory. It states that one can write down all the needed $\beta$-patterns for both the pQCD series and the PMC scale $\alpha_s$-expansion at any perturbative order by analyzing the running behavior of the coupling constant determined by the RGE; i.e., by following the $\beta$-pattern of Eq.(\ref{betaseries}). The PMC scales themselves will also have a perturbative expansion (in an exponential form) in order to achieve a consistent resummation of all $\beta$-terms into the coupling constant~\cite{kataev1}. More explicitly, by writing out the $\beta$-series, Eq.(\ref{betaseries}) can be rewritten as
\begin{widetext}
\begin{eqnarray}
a_{s}(\mu) &=& a_{s}(\mu_0)-\frac{1}{4} \beta_{0} \ln\left(\frac{\mu^2}{\mu_0^2}\right) a^{2}_{s}(\mu_0) +\frac{1}{4^2}\left[\beta^2_{0} \ln^2 \left(\frac{\mu^2}{\mu_0^2}\right) -\beta_{1} \ln\left(\frac{\mu^2}{\mu_0^2}\right) \right] a^{3}_{s}(\mu_0) + \nonumber\\
&& \frac{1}{4^3}\left[-\beta^3_{0} \ln^3 \left(\frac{\mu^2}{\mu_0^2}\right) +\frac{5}{2} \beta_{0}\beta_{1} \ln^2\left(\frac{\mu^2}{\mu_0^2}\right) -\beta_{2} \ln\left(\frac{\mu^2}{\mu_0^2}\right)\right] a^{4}_{s}(\mu_0) +{\cal O}(a^{5}_{s}) , \label{astrans}
\end{eqnarray}
\end{widetext}
where $a_s=\alpha_s/\pi$, $\mu_0$ stands for initial renormalization scale and the scheme parameter has been omitted for convenience.

Following the idea of PMC-BLM correspondence principle, we are ready to obtain the PMC scales via a systematic way. In general, by identifying the $n_f$-terms explicitly at each perturbative order, the pQCD prediction $\varrho_n$ for a physical observable $\varrho$ up to four-loop level can be rewritten as
\begin{widetext}
\begin{displaymath}
\varrho_{n} = r_0 \left[ a^p_s(\mu_0) + \sum^2_{i=1} A_i n_f^{i-1} a^{p+1}_s(\mu_0) + \sum^3_{i=1} B_i n_f^{i-1} a^{p+2}_s(\mu_0) + \sum^4_{i=1} C_i n_f^{i-1} a^{p+3}_s(\mu_0) \right ]
\end{displaymath}
\end{widetext}
where $r_0$ is scale-independent and is free from $a_s(\mu_0)$, $p\; (\geq1)$ stands for the leading-order $\alpha_s$ power. The PMC scales for $\varrho_{n}$ can be determined in a step-by-step way such that all those $n_f$-terms will be absorbed into the running coupling. That is, the PMC-I approach suggests that the QCD corrections are formed by a sequential one-loop and one-loop corrections, and one can inversely set the PMC scale for a $\alpha_s$ order by resuming all $n_f$-terms with highest power in all higher-order $\alpha_s$ terms into this particular $\alpha_s$ order. More specifically,
\begin{widetext}
\begin{itemize}
\item The first step is to set the PMC scale $Q_1$ at LO, which is derived by absorbing $A_2 n_f$, $B_3 n_f^2$ and $C_4 n_f^3$ into $a^p_s$:
\begin{equation}
\varrho^{\prime}_n = r_0\Big[ {a^p_s(Q_1)} + \widetilde{A}_{1} a^{p+1}_s(Q_1) + (\widetilde{B}_{1} + \widetilde{B}_{2}n_f) a^{p+2}_s(Q_1) + (\widetilde{C}_{1} +\widetilde{C}_{2}n_f +\widetilde{C}_{3}n_f^2) a^{p+3}_s(Q_1) \Big] . \label{first}
\end{equation}
\item The second step is to set the effective scale $Q_2$ at NLO, which is derived by absorbing $\widetilde{B}_2 n_f$ and $\widetilde{C}_3 n_f^2$ into $a^{p+1}_s$:
\begin{eqnarray}
\varrho^{\prime\prime}_n &=& r_0\Big[ a^p_s(Q_1) + {\widetilde{A}_{1} a^{p+1}_s(Q_2)} + \widetilde{\widetilde{B}}_{1} a^{p+2}_s(Q_2) + (\widetilde{\widetilde{C}}_{1} +\widetilde{\widetilde{C}}_{2}n_f) a^{p+3}_s(Q_2) \Big]\ , \label{second}
\end{eqnarray}
\item The final step is to set the effective scale $Q_3$ at N$^2$LO, which is derived by absorbing $\widetilde{\widetilde{C}}_2 n_f$ into $a^{p+2}_s$:
\begin{eqnarray}
\varrho^{\prime\prime\prime}_n &^=& r_0\Big[ a^p_s(Q_1) + \widetilde{A}_{1} a^{p+1}_s(Q_2) + {\widetilde{\widetilde{B}}_{1} a^{p+2}_s(Q_3)} + \widetilde{\widetilde{\widetilde{C}}}_{1} a^{p+3}_s(Q_3) \Big]\ . \label{third}
\end{eqnarray}
\end{itemize}
\end{widetext}
When performing the shifts $\mu_0\to Q_1$, $Q_1 \to Q_2$ and $Q_2\to Q_3$, we eliminate the $n_f$-terms associated with the corresponding $\beta$-terms completely. Those step-by-step coefficients can be calculated by sequentially setting $\varrho^{\prime}_n=\varrho_n$, $\varrho^{\prime\prime}_n =\varrho^{\prime}_n$ and $\varrho^{\prime\prime\prime}_n =\varrho^{\prime\prime}_n$, which can be found in Ref.\cite{pmc2}. At the same time, we also have to modify the coefficients such that the final ones are conformal. We have no $\beta$-terms to set the PMC scale for $a^{p+3}_s$, so in practice we will set its value as the determined one-order-lower PMC scale $Q_3$. Thus, there is residual scale dependence due to those unknown $\beta$-terms from higher-order QCD prediction.

The PMC scales up to N$^2$LO can be written as
\begin{widetext}
\begin{eqnarray}
\ln\frac{Q^{2}_1}{\mu^2_0} &=& \ln \frac{Q^{2}_{1,0}}{\mu_0^2} + \frac{x \beta_0}{4} \ln \frac{Q^{2}_{1,0}}{\mu_0^2} a_s(\mu_0) + \frac{y}{16}\left(\beta^2_0 \ln^2 \frac{Q^{2}_{1,0}}{\mu_0^2} -\beta_1 \ln\frac{Q^{2}_{1,0}}{\mu_0^2}\right) a^2_s(\mu_0) + {\cal O}(a^3_s) \\
\ln\frac{Q^{2}_{2}}{Q^{2}_{1}} &=& \ln\frac{Q^{2}_{2,0}}{Q^{2}_{1}} + \frac{z \beta_0}{4} \ln \frac{Q^{2}_{2,0}}{Q^{2}_{1}} a_s(\mu_0)+ {\cal O}(a^2_s) \\
\ln\frac{Q^{2}_{3}}{Q^{2}_{2}} &=& \ln \frac{Q^{2}_{3,0}}{Q^{2}_{2}}+ {\cal O}(a_s)
\end{eqnarray}
\end{widetext}
where the scales $Q_{1,0}$, $Q_{2,0}$ and $Q_{3,0}$ are determined so as to eliminate $A_2 n_f$, $\widetilde{B}_2 n_f$ and $\widetilde{\widetilde{C}}_2 n_f$-terms completely, the parameters $x$ and $z$ are used to eliminate the $B_3 n_f^2$ and the $\widetilde{C}_3 n_f^2$ terms respectively, and the parameter $y$ is used to eliminate the $C_4 n_f^3$-term. It is found that
\begin{displaymath}
\ln \frac{Q^{2}_{1,0}}{\mu_0^{2}} = \frac{6A_2}{p},
\ln \frac{Q^{2}_{2,0}}{Q_1^{2}} = \frac{6\widetilde{B}_2}{(p+1)\widetilde{A}_1},
\ln \frac{Q^{2}_{3,0}}{Q_2^{2}} = \frac{6\widetilde{\widetilde{C}}_2} {(p+2)\widetilde{\widetilde{B}}_1}
\end{displaymath}
and
\begin{eqnarray}
x &=& \frac{3(p+1)A_2^2 -6 p B_3}{p A_2} \\
y &=& \frac{(p+1)(2p+1)A_2^3 -6p(p+1)A_2 B_3 +6p^2 C_4}{p A^2_2} \\
z &=& \frac{3(p+2)\widetilde{B}_2^2 -6(p+1)\widetilde{A}_1 \widetilde{C}_3}{(p+1) \widetilde{A}_1 \widetilde{B}_2}
\end{eqnarray}

\subsection{PMC-II: Achieving the goal of PMC via $R_\delta$-scheme}

The PMC-I approach provides a way to set the PMC scales for any scheme, such as the ${\rm MS}$-scheme~\cite{MS1,MS2}, the $\overline{\rm MS}$ scheme~\cite{MSbar} and the ${\rm MOM}$-scheme~\cite{MOM}. As for the dimensional renormalization schemes similar to the ${\rm MS}$-scheme and $\overline{\rm MS}$-scheme, we can adopt a more convenient approach for setting the PMC scales~\cite{BMW,BMW2}. For convenience, we call it as the PMC-II approach.

The starting point of the PMC-II approach is to introduce an arbitrary dimensional renormalization scheme, the $R_\delta$-scheme. In the $R_\delta$-scheme, an arbitrary constant $-\delta$ is subtracted in addition to the standard subtraction $\ln 4 \pi - \gamma_E$ for the $\overline{\rm MS}$-scheme. This amounts to redefining the renormalization scale by an exponential factor, $\mu_\delta = \mu_{\overline{\rm MS}} \exp(\delta/2)$. The $\delta$-subtraction thus defines an infinite set of new renormalization schemes. All $R_\delta$-schemes are connected to each other by a scale-displacement; thus the $\beta$-function of the strong QCD coupling constant $\alpha = \alpha_s/(4\pi)$ is the same as usual $\overline{\rm MS}$ one, i.e.
\begin{equation}
\mu_\delta^2\frac{d \alpha}{d\mu_\delta^2} = \beta(\alpha) = - \alpha(\mu_\delta)^2 \sum_{i=0}^\infty \beta_i \alpha(\mu_\delta)^i \ .
\end{equation}

In contrast to the idea of loop-by-loop determination for the PMC-I approach, the PMC-II approach allows all PMC scales to be simultaneously determined. This makes the PMC scale-setting transparent and straightforward. In practice the PMC-I and PMC-II methods may lead to differences in the predictions for the individual PMC-scales, although we shall show they are equivalent for the final predictions.

Under the PMC-II approach, at each perturbative order, in analogy to Eq.(\ref{astrans}), the running behavior of the coupling constant is controlled by the displacement relation between couplings in any $R_\delta$-scheme
\begin{equation} \label{displace}
\alpha(\mu_0) = \alpha(\mu_\delta) + \sum_{n=1}^\infty \frac{1}{n!} { \frac{{\rm d}^n \alpha(\mu)}{({\rm d} \ln \mu^2)^n}\left|_{\mu=\mu_\delta}\right. (-\delta)^n}
\end{equation}
where $\ln\mu^2_0/\mu^2_\delta=-\delta$. Eq.(\ref{displace}) indicates the $\{\beta_i\}$-terms that pertain to a specific perturbative order. By collecting up all those $\{\beta_i\}$-terms for the same order, one can obtain the general pattern of nonconformal $\{\beta_i\}$-terms at each perturbative order. That is, by using $R_\delta$-scheme, we can rewrite the pQCD prediction of a physical observable $(\varrho)$ up to $\alpha^4$ as~\cite{BMW,BMW2}
\begin{widetext}
\begin{eqnarray}
\varrho_\delta(Q) &=& r_0 + r_1 \alpha_1(\mu_1) + \left[r_2 + \beta_0 r_1 \delta_1\right] \alpha^2_2(\mu_2) + \left[r_3 +\beta _1 r_1 \delta_1+ 2 \beta _0 r_2\delta_2+ \beta _0^2 r_1 \delta_1^2 \right] \alpha^3_3(\mu_3) \nonumber\\
&& + \left[r_4 +\beta _2 r_1\delta_1 +2 \beta _1 r_2\delta_2 +3 \beta _0 r_3\delta_3 +3 \beta _0^2 r_2\delta_2^2 +\beta _0^3 r_1 \delta_1 ^3 +\frac{5}{2} \beta _1 \beta _0 r_1\delta_1^2 \right] \alpha^4_4(\mu_4), \label{rdeltas}
\end{eqnarray}
\end{widetext}
where $\mu_i = Q e^{\delta_i/2}$, the initial scale $\mu_0$ is for simplicity set to be $Q$ at which the observable is measured. To best illuminate the method, we have put an artificial index on each $\alpha$ and $\delta$ to keep track of which coupling each $\delta$-term is associated with. Eq.(\ref{rdeltas}) also reveals a special degeneracy of the terms in the perturbative coefficients at different orders such that one can achieve an one-to-one correspondence between $\beta$-terms and $n_f$-terms as PMC-I does. Then, the QCD prediction $\varrho_n$ of a physical observable $\varrho$ up to four-loop level can be expressed as~\cite{BMW,BMW2}
\begin{widetext}
\begin{eqnarray}
\varrho_{n}(Q) &= & r_{0,0} + r_{1,0} \alpha(Q) + \left[r_{2,0} + \beta_0 r_{2,1} \right] \alpha^2(Q) + \left[r_{3,0} + \beta_1 r_{2,1} + 2 \beta_0 r_{3,1} + \beta _0^2 r_{3,2} \right] \alpha^3(Q) \nonumber\\
&& +[r_{4,0} + \beta_2 r_{2,1} + 2\beta_1 r_{3,1} + \frac{5}{2} \beta_1 \beta_0 r_{3,2} +3\beta_0 r_{4,1} + 3 \beta_0^2 r_{4,2} + \beta_0^3 r_{4,3} ] \alpha^4(Q), \label{betapattern}
\end{eqnarray}
\end{widetext}
where $Q$ stands for the scale at which it is measured, the $r_{i,0}$ are the conformal parts of the perturbative coefficients. Here for convenience we have set the initial scale $\mu_0=Q$.

Using PMC-II, it can be shown that the order $\alpha^k(Q)$ coupling must be resummed into the effective coupling $\alpha^k(Q_k)$, given by:
\begin{widetext}
\begin{eqnarray}
r_{1,0} \alpha(Q_1) &=& r_{1,0} \alpha(Q) - \beta(\alpha) r_{2,1} + \frac{1}{2} \beta(\alpha) \frac{\partial \beta}{\partial \alpha} r_{3,2} + \cdots + \frac{(-1)^n}{n!} \frac{{\rm d}^{n-1}\beta}{({\rm d} \ln\mu^2)^{n-1}} r_{n+1,n} \ , \nonumber \\
& \hspace{2mm} \vdots &\nonumber \\
r_{k,0} \alpha^{k}(Q_k) &=& r_{k,0} \alpha^k(Q) + r_{k,0}\ k \ \alpha^{k-1}(Q) \beta(\alpha) \left \{ S_{k,1} +\Delta_k^{(1)}(\alpha) S_{k,2} + \cdots + \Delta_k^{(n-1)}(\alpha) S_{k,n} \right \} \ ,
\label{korder}
\end{eqnarray}
which defines the PMC scales $Q_k$, and where we have introduced
\begin{eqnarray}
S_{k,j}=(-1)^j \frac{r_{k+j,j}}{r_{k,0}}\ ,\\
\Delta_k^{(1)}(\alpha) &=& \frac{1}{2} \left [ \frac{\partial \beta}{\partial \alpha} + (k-1) \frac{\beta}{\alpha}\right] \ , \\
\Delta_k^{(2)}(\alpha) & =& \frac{1}{3!}\left [ \beta \frac{\partial^2 \beta}{\partial \alpha^2} + \left( \frac{\partial\beta}{\partial \alpha} \right )^2 + 3(k-1) \frac{\beta}{\alpha}\frac{\partial\beta}{\partial \alpha} + (k-1)(k-2) \frac{\beta^2}{\alpha^2}\right ] \ . \\
\hspace{2mm} \vdots \nonumber
\end{eqnarray}
\end{widetext}

Eq.(\ref{korder}) is systematically derived by replacing the $\ln^j Q_1^2/Q^2$ by $S_{k,j}$ in the logarithmic expansion of $\alpha^k(Q_k)$ up to the highest known $S_{k,n}$-coefficient in pQCD. The resummation can be performed iteratively using the RG equation for $\alpha$ and leads to the effective scales for an ${\rm N^3LO}$ prediction:
\begin{widetext}
\begin{eqnarray} \label{exactscales}
\ln \frac{Q_{k}^2}{Q^2} &=& \frac{S_{k,1} + \Delta_k^{(1)}(\alpha) S_{k,2} +\Delta_k^{(2)}(\alpha) S_{k,3}}{1+ \Delta_k^{(1)}(\alpha) S_{k,1} + \left({\Delta_k^{(1)}(\alpha)}\right)^2 (S_{k,2} - S_{k,1}^2) + \Delta_k^{(2)}(\alpha) S_{k,1}^2 } \ .
\end{eqnarray}
\end{widetext}

After setting the PMC scales $Q_i$, the final pQCD prediction for $\varrho_n$ up to four-loop level then reads
\begin{equation}
\varrho_n(Q) = r_{0,0} + \sum_{i=1}^{4} r_{i,0} \alpha^{i}(Q_i) \ .
\end{equation}
Here $Q_4$ remains unknown and causes the residual scale dependence, since it requires the knowledge of $r_{5,1}$ in the coefficient of $\alpha^5$. One can as a convention set its value as the initial renormalization scale, or more reasonably, set its value as the determined one-order-lower PMC scale $Q_3$. Since the $\delta$ and $\beta$ terms are resummed into the running coupling, the PMC-II prediction automatically satisfies the RGI principle. In principle, one can use measurements of $\alpha_s$ at $Q=M_z$ to determine a value for the QCD coupling in $R_\delta$ scheme including $\delta =0$. Thus the PMC-II predictions are scheme independent at any finite order.

\section{PMS and Local RGI} \label{sec:pms}

The PMS introduces local RGI to set the renormalization scale: if an estimate depends on some ``unphysical" parameters, then their values should be chosen so as to minimize the sensitivity of the estimate to small variations of those parameters~\cite{PMS1,PMS2,PMS3,PMS4}.

As an illustration of the PMS, we expand the N$^{n}$LO approximant $\varrho_n(Q)$ as
\begin{eqnarray} \label{pmsstartpoint}
\varrho_n(Q) = a_s(\mu) \left(1 + \sum_{i=1}^{n}{\cal C}_i(\mu,Q) a_s^i(\mu) \right) ,
\end{eqnarray}
where $Q$ is the scale at which $\varrho$ is measured and $a_s=\alpha_s/\pi$. The local RGI indicates that
\begin{widetext}
\begin{eqnarray}
\frac{\partial \varrho_n}{\partial \tau}&=&\left(\left.\frac{\partial }{\partial \tau}\right|_{a_s} +\beta (a_s) \frac{\partial}{\partial (a_s/4)} \right)\varrho_n=0 , \label{pmsstart1} \\
\frac{\partial \varrho_n}{\partial \beta_m}&=&\left(\left.\frac{\partial }{\partial \beta_m}\right|_{a_s} -\beta (a_s)\int_0^{a_s/4} d\left(\frac{a_s^\prime}{4}\right) \frac{(a_s^\prime/4)^{m+2}}{\left[\beta (a_s^\prime)\right]^2} \frac{\partial} {\partial (a_s/4)} \right)\varrho_n=0,\; (m=2,3,...) \label{pmsstart2}
\end{eqnarray}
\end{widetext}
where $\tau =\ln(\mu^2/\tilde\Lambda^2_{\rm{QCD}})$. The QCD parameter $\tilde\Lambda_{\rm{QCD}}$ is related to the conventional $\Lambda_{\rm{QCD}}^{\overline{\rm{MS}}}$ through the relation~\cite{PMS1}
\begin{eqnarray}
\tilde\Lambda_{\rm{QCD}}=\left(\frac{\beta_1}{\beta_0^2}\right)^{-\beta_1/2\beta_0^2} \Lambda_{\rm{QCD}}^{\overline{\rm{MS}}}. \label{lambdarelation}
\end{eqnarray}

Substituting Eq.(\ref{pmsstartpoint}) into Eqs.(\ref{pmsstart1},\ref{pmsstart2}) and equating powers of $a_s$, one finds ${\cal C}_1$ depends on $\tau$ only, while ${\cal C}_2$ depends on $\tau$ and $\beta_2$, and etc. More explicitly, we have
\begin{eqnarray}
&&{\partial {\cal C}_1 \over \partial \tau}= {1 \over 4}\beta_0 \qquad \qquad \qquad
{\partial {\cal C}_1 \over \partial \beta_2} = 0 \\
&&{\partial {\cal C}_2 \over \partial\tau}={1 \over 2}\beta_0 {\cal C}_1
+ {1\over 16}\beta_1 \quad \ \
{\partial {\cal C}_2 \over \partial \beta_2} = - {1 \over 16}{1 \over \beta_0} \label{3loopd} \\
&& \quad\quad\quad\quad\quad\quad\quad \cdots\cdots \nonumber
\end{eqnarray}
These differential equations show that the perturbative coefficient ${\cal C}_n$ is in general a function of $\tau$ and the scheme parameters $\beta_2$, $\beta_3$, $\cdots$, plus a local RG invariant integration constant $\rho_n$. To be locally RG invariant means that a coefficient is independent of $\tau$ and scheme parameters $\{\beta_i\}$. As the key point of PMS, following the condition (\ref{pmsconst}), those local RG invariants shall be determined in an order-by-order way, i.e. once they have been determined, they should not be changed by any higher-order corrections. For example, at ${\rm N^{2}LO}$ level, we need to introduce two local RG invariants
\begin{eqnarray}
\rho_1 &=& {1\over 4} \beta_0 \tau -{\cal C}_1 ,\,\,\label{rho1} \\
\rho_2 &=& {\cal C}_2 - \left ( {\cal C}_1+{1 \over 8}{\beta_1 \over \beta_0} \right )^2 + {1 \over 16}{ \beta_2 \over \beta_0}, \label{rho2}
\end{eqnarray}
and for ${\rm N^{3}LO}$ level, $\rho_1$ and $\rho_2$ are fixed and we need to introduce an extra local RG invariant
\begin{eqnarray}
\rho_3 &=& \frac{{{\beta _3}}}{{64{\beta _0}}} + \frac{{{\beta _1}{\cal C}_1^2}}{{4{\beta _0}}} - \frac{{{\beta _2}{{\cal C}_1}}}{{8{\beta _0}}} + 4{\cal C}_1^3 - 6{{\cal C}_2}{{\cal C}_1} + 2{{\cal C}_3} . \label{rho3}
\end{eqnarray}

One can obtain an expression for $\tau$ from the scale equation (\ref{scale}) via proper parameter transformation,
\begin{widetext}
\begin{eqnarray}
\tau &=& \int_{a_s/4}^\infty d \left( \frac{x}{4} \right) \frac{-1}{\beta^{(m)}(x)} = \frac{4}{\beta_0 a_s} + \frac{\beta_1}{\beta_0^2}\ln \left| \frac{\beta_1 a_s} {\beta_1 a_s + 4\beta_0}\right| + \int_0^{a_s/4} d \left( \frac{x}{4} \right) \left(\frac{1}{\beta^{(m)}(x)} -\frac{1} {\beta^{(2)}(x)} \right), \label{betaintegrate}
\end{eqnarray}
\end{widetext}
where $\beta^{(m)}$ stands for the cut $\beta$-function up to $a_s^{m+2}$. This equation can be solved numerically or analytically.

With this basis, we can derive the optimal behavior for $\varrho_n$. We first consider a NLO approximant
\begin{eqnarray}
\varrho_1 &=& a_s+{\cal C}_1 a_s^2 =\tilde{a}_s + \tilde{\cal C}_{1} \tilde{a}_s^2 ,
\end{eqnarray}
where the first equality is the estimate assuming any renormalization scheme (usually the $\overline{\rm MS}$-scheme), and the second equality one stands for the optimized prediction after applying PMS. The approximant $\varrho_1$ depends on scheme and scale only through the variable $\tau$. From Eq.(\ref{betaintegrate}), we obtain
\begin{eqnarray}
\tau=\frac{4}{\beta_0 a_s}+\frac{\beta_1}{\beta_0^2}\ln \left| \frac{\beta_1 a_s}{\beta_1 a_s +4\beta_0}\right| . \label{NLOtau}
\end{eqnarray}

From Eq.(\ref{pmsstart1}), we obtain the local RGI equation
\begin{eqnarray}
\beta_0-(1+2\tilde {\cal C}_1 \tilde a_s) \left(\beta_0+\beta_1 \frac{\tilde a_s}{4}\right) =0, \label{NLOPMS}
\end{eqnarray}
which leads to
\begin{eqnarray}
\tilde {\cal C}_1=-\frac{\beta_1}{2\beta_1 \tilde a_s+8\beta_0}. \label{NLOr1}
\end{eqnarray}
Together with Eqs.(\ref{rho1},\ref{NLOtau},\ref{NLOr1}), we finally obtain
\begin{eqnarray}
\frac{1}{\tilde a_s}+\frac{\beta_1}{2\beta_1 \tilde a_s+8\beta_0} +\frac{\beta_1}{4\beta_0} \ln\left| \frac{\beta_1\tilde a_s}{\beta_1 \tilde a_s+4\beta_0}\right| = \rho_1 .
\end{eqnarray}
One can numerically solve this equation to obtain $\tilde{a}_s$, find out $\tilde {\cal C}_1$ and $\tau$, and get the optimized estimate for $\varrho_1$. The above procedures can be extended to any order. Specifically, we adopt the ${\rm N^{3}LO}$ approximant as an explanation of how to deal with it in higher orders, which can be directly adopted to deal with the pQCD prediction for $R(e^+ e^-)$ and $H\to b\bar{b}$ up to four-loop level.

The ${\rm N^{3}LO}$ approximant can be written as,
\begin{eqnarray} \label{startpoint}
\varrho_3 &=& a_s + {\cal C}_1 a_s^2 + {\cal C}_2 a_s^3 + {\cal C}_3 a_s^4 \nonumber \\
   &=& \tilde{a}_s + \tilde{\cal C}_1 \tilde{a}_s^2 + \tilde{\cal C}_2 \tilde{a}_s^3 + \tilde{\cal C}_3 \tilde{a}_s^4 .
\end{eqnarray}
At present, the scheme and scale dependence of $\varrho_3$ is controlled by $\tau$, $\beta_2$ and $\beta_3$. The local RGI equations (\ref{pmsstart1},\ref{pmsstart2}) can be written as:
\begin{widetext}
\begin{eqnarray}
\tilde a_s^3 \tilde \beta_3 \tilde {\cal C}_3+16 \tilde a_s^2 \tilde \beta_2\tilde{\cal C}_3+3\tilde a_s^2\tilde\beta_3\tilde{\cal C}_2+64\tilde a_s\beta_1\tilde{\cal C}_3+12\tilde a_s\tilde\beta_2\tilde{\cal C}_2+2\tilde a_s\tilde\beta_3\tilde{\cal C}_1+\tilde\beta_3+256\beta_0\tilde{\cal C}_3+48\beta_1\tilde{\cal C}_2+8\tilde\beta_2\tilde{\cal C}_1=0, \label{pmsnnnlo1}\\
\beta_0(3\tilde a_s\tilde\beta_3+8\tilde\beta_2)(4\tilde{\cal C}_3\tilde a_s^3+3\tilde{\cal C}_2\tilde a_s^2+2\tilde{\cal C}_1\tilde a_s+1)-\tilde a_s\beta_1\tilde\beta_2(4\tilde{\cal C}_3\tilde a_s^3+3\tilde {\cal C}_2\tilde a_s^2+2\tilde{\cal C}_1\tilde a_s+1)+384\beta_0^2(4\tilde{\cal C}_3\tilde a_s+3\tilde{\cal C}_2)=0, \label{pmsnnnlo2}\\
\beta_1^2\tilde a_s(4\tilde{\cal C}_3\tilde a_s^3+3\tilde{\cal C}_2\tilde a_s^2+2\tilde{\cal C}_1\tilde a_s+1)+96\beta_0^2(4\tilde {\cal C}_3\tilde a_s^2+3\tilde{\cal C}_2\tilde a_s+2\tilde{\cal C}_1)-8\beta_0\beta_1(4\tilde{\cal C}_3\tilde a_s^3+3\tilde{\cal C}_2\tilde a_s^2+2\tilde {\cal C}_1\tilde a_s+1)=0 , \label{pmsnnnlo3}
\end{eqnarray}
\end{widetext}
where $\tilde\beta_2$ and $\tilde\beta_3$ are $\beta$-functions under the optimized scheme. Together with the equations (\ref{rho1},\ref{rho2},\ref{rho3}) for the RG-invariants $\rho_{1,2,3}$ and the scale running equation (\ref{betaintegrate}), we have to solve seven equations simultaneously. Note tat all parameters in these formulae should be changed to tilde ones accordingly. For this purpose, we adopt the so-called `spiraling' method~\cite{pmsfixedpoint} to solve them iteratively and numerically. The main procedure is
\begin{enumerate}
\item Choose an initial value for $\tilde{a}_s$.
\item Set the initial values for $\tilde\beta_2$ and $\tilde\beta_3$ to be $\beta_2$ and $\beta_3$ for the first iteration or as the values determined from last iteration. Solve Eqs.(\ref{pmsnnnlo1},\ref{pmsnnnlo2},\ref{pmsnnnlo3}) for $\tilde{\cal C}_1$, $\tilde{\cal C}_2$ and $\tilde{\cal C}_3$.
\item Apply the calculated $\tilde{\cal C}_1$, $\tilde{\cal C}_2$ and $\tilde{\cal C}_3$ into the equations (\ref{rho1},\ref{rho2},\ref{rho3},\ref{betaintegrate}) for $\tilde{a}_s$, $\tau$, $\tilde\beta_2$ and $\tilde\beta_3$.
\item Iterate from second step until the results for $\varrho_3$ converge to an acceptable prediction.
\end{enumerate}

\section{Comparative Studies of PMC and PMS} \label{sec:comp}

As indicated by Eq.(\ref{rho1}), after applying PMS, the NLO coefficient is obtained by shifting the $\beta_0$-term into the running coupling. The PMC and PMS predictions are different even at the NLO level, since the PMC and PMS scales are different. Taking three-jet production in $e^+e^-$-annihilation as an example, it has been observed that the PMS scale cannot yield the correct physical behavior for the normalization scale for $e^+ e^- \to q \bar q g$, since the renormalization scale rises anomalously without bound for small jet energy~\cite{Kramer1,Kramer2}. In contrast, the PMC scale has the correct behavior~\cite{pmc1}.

For N$^2$LO and even higher-order calculations, the conditions are much more complicated. In the following subsections, we present two explicit examples for a detailed comparison of PMS and PMC up to four-loop level.

\subsection{$R_{e^+e^-}$ up to four-loop level}
\label{sec:Ree}

The electron-positron annihilation into hadrons provides one of the most precise platforms for testing the $\alpha_s$ behavior. The usual $R$-ratio is defined as
\begin{eqnarray}
R_{e^+e^-}(Q)&=&\frac{\sigma\left(e^+e^-\rightarrow {\rm hadrons} \right)}{\sigma\left(e^+e^-\rightarrow \mu^+\mu^-\right)}\nonumber\\
&=& 3\sum_q e_q^2\left[1+R(Q)\right], \label{Re+e-}
\end{eqnarray}
where $Q$ stands for the energy at which it is measured. Theoretically, the pQCD prediction for $R$ up to (n+1)-loop correction $R_n$ can be written as
\begin{eqnarray}
R_n(Q)=\sum_{i=0}^{n} {\cal C}_{i}(Q,\mu_0) a_s^{i+1}(\mu_0), \label{R(Q)}
\end{eqnarray}
where $a_s=\alpha_s/\pi$. At present, the pQCD prediction for $R(Q)$ has been calculated within the $\overline{\rm MS}$-scheme up to four-loop level~\cite{Ralphas1,Ralphas2}. In order to apply the PMC scale setting correctly, i.e. only those $n_f$-terms that rightly determine the running behavior of the coupling constant should be resummed into the coupling constant~\cite{BMW,BMW2,Wu:2013ei}, we adopt the $R(Q)$ expression derived by analytically continuing the Adler function $D$ into the time-like region~\cite{Ralphas3,Ralphas4}, where $D(Q^2) = \gamma(a) - \beta(a) \frac{d}{da_s} \Pi(Q^2,a_s)$ where $\gamma$ is the anomalous dimension of the vector field and $\Pi$ the vacuum polarization function.

\subsubsection{properties of $R_n$}

\begin{figure}[tb]
\centering
\includegraphics[width=0.6\textwidth]{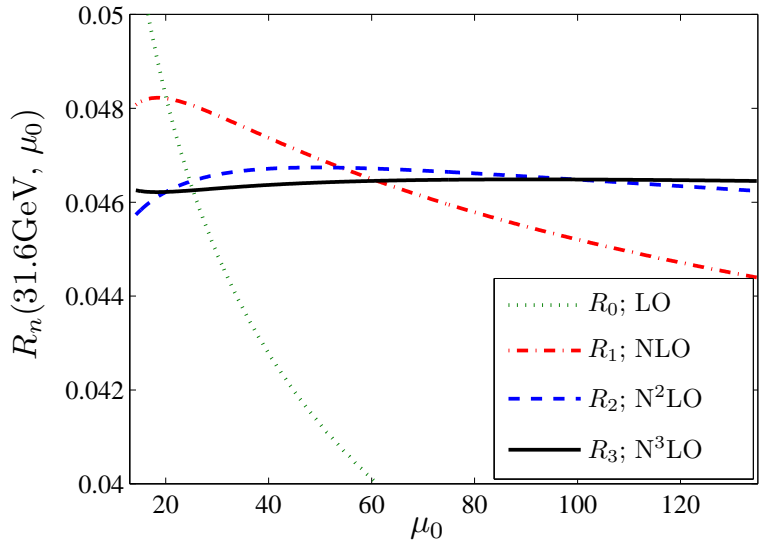}
\caption{The pQCD prediction $R_{n}(Q=31.6 {\rm GeV},\mu_{0})$ up to four-loop level versus the initial scale $\mu_0$. In conventional scale setting the $\mu_0$ dependence is used as a measure of the renormalization-scale ``uncertainty", since the initial renormalization-scale and scheme dependence is conventionally left untreated. The dotted, the dash-dot, the dashed and the solid lines are for $R_0$, $R_1$, $R_2$ and $R_3$, respectively. } \label{Rsmu1}
\end{figure}

To do the numerical calculation for $R_3$, we will adopt $\Lambda_{\overline{\rm{MS}}}^{(n_f=5)} =213$ MeV~\cite{pmc2,bethke}, which is determined from $R_{e^+ e^-}$ by using the four-loop $\alpha_s$ running with $\alpha^{\overline{\rm MS}}_s(M_z)=0.1184$~\cite{pdg}. The $\Lambda_{\overline{\rm{MS}}}^{(n_f=5)}$ values for $R_i$ with $i<3$ can be determined by using the $(i+1)_{\rm th}$-loop $\alpha_s$ running via a similar way.

We start from the scale dependence of $R_{n}$ using conventional scale setting. Under such scale setting, the scale dependence from $a_s$ and ${\cal C}_i$ do not exactly cancel at any finite order, and $R_n$ depends on both $Q$ and $\mu_{0}$. The results of $R_{n}$ up to four-loop level are presented in FIG. \ref{Rsmu1}, where we set $Q=31.6$ GeV~\cite{experimentRee}. It shows the one-loop and two-loop predictions $R_0$ and $R_1$ strongly depend on $\mu_{0}$. When more loops have been taken into consideration, one obtains a weaker scale dependence. This agrees with the conventional wisdom that by finishing a higher-and-higher order calculation, one can get a desirable scale-invariant estimate.

More explicitly, we find the four-loop prediction for $R_3$ depends slightly on the scale choice: by varying $\mu_{0}\in[Q/2,2Q]$, we have $\left.\frac{\Delta R_3(Q,\mu_0)} {R_3(Q,Q)}\right|_{\rm Conv.}=\left(^{+0.4\%}_{-0.2\%}\right)$ for the conventional scale setting; The residual scale dependence for PMC due to unknown higher order $\{\beta_i\}$-terms is $\left.\frac{\Delta R_3(Q,\mu_0)} {R_3(Q,Q)}\right|_{\rm PMC-I}=\left(^{+0.2\%}_{-0.0\%}\right)$ and $\left.\frac{\Delta R_3(Q,\mu_0)} {R_3(Q,Q)}\right|_{\rm PMC-II}=\left(^{+0.2\%}_{-0.2\%}\right)$. Here $\Delta R_3(Q,\mu_0)=R_3(Q,\mu_0)-R_3(Q,Q)$. As for PMS, its prediction only depends on at what scale it is measured, since the initial scale dependence has been absorbed into the local RG invariants $\rho_i$.

\begin{table}[tb]
\centering
\begin{tabular}[b]{cccccccc}
\hline
 & ~~$R_1$~~ & ~~$R_2$~~ & ~~$R_3$~~ & ~~$\kappa_1$~~ & ~~$\kappa_2$~~ & ~~$\kappa_3$~~ \\ \hline
~~Conv.~~ & 0.04777 & 0.04662 & 0.04631 & $7.35\%$ & $-2.41\%$ & $-0.66\%$ \\
PMC-I & 0.04759 & 0.04645 & 0.04627 & $6.94\%$ & $-2.40\%$ & $-0.39\%$ \\
PMC-II & 0.04759 & 0.04663 & 0.04631 & $6.94\%$ & $-2.02\%$ & $-0.69\%$ \\
PMS  & 0.04880 & 0.04640 & 0.04633 & $9.66\%$ & $-4.92\%$ & $-0.15\%$ \\
\hline
\end{tabular}
\caption{Numerical results for $R_n$ and $\kappa_n$ with various QCD loop corrections under the conventional scale setting (Conv.), PMC-I, PMC-II and PMS, respectively. The value of $R_0=0.04450$ is the same for all scale settings. $Q=31.6$ GeV and $\mu_0=Q$. } \label{table1}
\end{table}

Numerical results for $R_n$ with various loop corrections are presented in Table \ref{table1}, where we have set $Q=31.6$ GeV and $\mu_0=Q$ for all scale settings. At the one-loop level, we have no information to set its scale, so all the scales are fixed to be $\mu_{0}(=Q)$ and we obtain $R_0=0.04450$ for all scale settings. To be consistent, as an estimate of $R_n$ we shall adopt $(n+1)$-loop $\alpha_s$-running behavior to do the calculation. To show how the theoretical prediction changes as more-and-more loop corrections are included, we define a ratio
\begin{equation}
\kappa_{n}=\frac{R_n-R_{n-1}}{R_{n-1}}, \label{kappa0}
\end{equation}
where $n=1,2,3$ respectively. This ratio shows how the (`known') lower-order estimate could be varied by a (`newly') available higher-order correction. As a comparison, we also present the results for PMC and PMS. Table \ref{table1} shows that all those scale-setting methods have a satisfactory steady behavior for $R_n$ when more loop corrections are included. At the four-loop level, the absolute values of $\kappa_3$ under various scale settings are less than $1\%$, and the $R_3$ under various scale setting are almost the same. Following the trends of the predictions, one may expect that the physical value $R$ could be $\sim 0.0463$.

We note that because of the slow scale dependence as shown by FIG. \ref{Rsmu1}, a guess of $\mu_{0}$ could lead to a value close to the experimental result for $R_{e^+e^-}$ using conventional scale setting; however, this may not be the correct answer at any fixed-order for general process. If a process does not converge quick enough, one has to use a more-and-more complex loop calculation to achieve the same precision goal as PMC and PMS. The problem is compounded by the $n!$ growth of the renormalon terms.

\begin{figure}[tb]
\centering
\includegraphics[width=0.6 \textwidth]{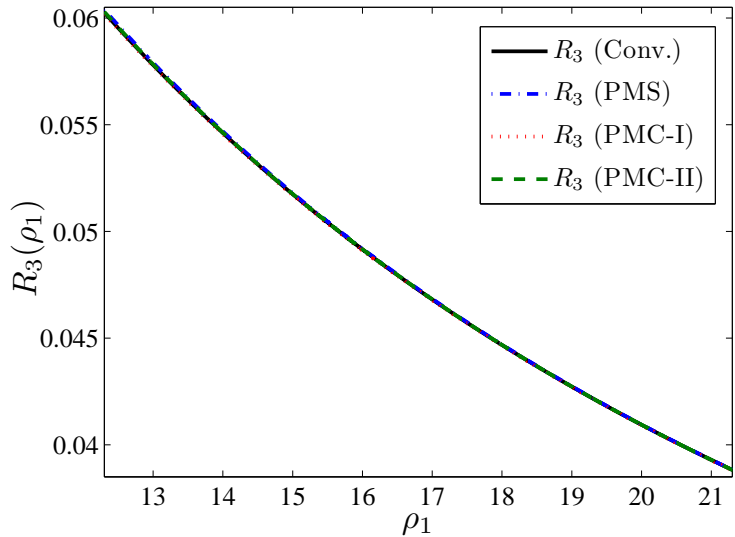}
\caption{Results for the four-loop estimate $R_3(\rho_1)$ versus $\rho_1\in [12,21]$ under various scale settings. The solid, the dotted, the dashed and the dash-dot lines are for conventional scale setting (Conv.), PMC-I, PMC-II and PMS, respectively. The conventional result depends on the initial scale.
All curves are almost coincide with each other. }\label{fit}
\end{figure}

\begin{table}[htb]
\centering
\begin{tabular}{ c c c c c c c c}
\hline
 & ~~Conv.~~ & ~~PMC-I~~ & ~~PMC-II~~ & ~~PMS~~ \\ \hline
~~$\Lambda^{(5)}_{20}$[MeV]~~ &$435^{+292}_{-206}$ & $437^{+294}_{-207}$ & $434^{+290}_{-206}$ & $431^{+286}_{-203}$\\
$\Lambda^{(5)}_{31.6}$[MeV] &$417^{+220}_{-166}$ & $419^{+221}_{-167}$ & $416^{+219}_{-166}$ & $414^{+217}_{-164}$ \\
$\Lambda^{(5)}_{\rm QCD}$[MeV] &$424\pm104$ & $426\pm105$ & $423\pm104$ & $421\pm103$ \\
\hline
\end{tabular}
\caption{Predictions of $\Lambda^{(5)}_{\rm QCD}$ from a comparison of four-loop estimates $R_3$ under various scale settings with two measurements $R(31.6\rm{GeV})$ and $R(20\rm{GeV})$ done by Ref.\cite{experimentRee}. The last line stands for the weighted average. } \label{table3}
\end{table}

We next show how the four-loop prediction for $R_3$ depends on the $e^+ e^-$ collision energy $Q$. The results for $R_3(\rho_1)$ under different scale settings are shown in FIG. \ref{fit}. In drawing the curves, we use $\rho_1$ (defined in Eq.(\ref{rho1})) instead of $Q$ as the argument of $R_3$ to avoid the uncertainty from the choice of $\Lambda_{\rm QCD}$~\cite{PMS5,Ma:2014oba}. For the chosen energy range ($Q>9{\rm GeV}$), we have $\rho_1\in(12,21)$. FIG. \ref{fit} shows that the four-loop estimate for $R_3(\rho_1)$ under various scale settings almost coincide with each other, which is consistent with Table \ref{table1}. Conversely, one can use the curves in FIG. \ref{fit} to determine the value of $\Lambda_{\rm{QCD}}$ by fitting them to the known experimental data~\cite{N$^2$LOanalysis}. For example, the values of $\Lambda^{(5)}_{\rm QCD}$ determined by taking the experimental measurements $R(Q=31.6\rm{GeV})=0.0527\pm0.0050$ and $R(Q=20\rm{GeV})=0.0587 \pm 0.0075$~\cite{experimentRee} are presented in Table \ref{table3}. Using the weighted average $\Lambda^{(5)}_{\rm QCD}$ listed in the last line of Table \ref{table3}, we predict
\begin{equation}
\alpha_s^{\overline{\rm{MS}}}(M_Z) = 0.132^{+0.005}_{-0.006},
\end{equation}
where different scale settings result in almost the same prediction for $\alpha_s^{\overline{\rm{MS}}}(M_Z)$. Even though the above value is slightly larger than the world average shown in Ref.\cite{pdg}, they agree well with the values obtained from the $e^+ e^-$ collider, {\it i.e.}, $\alpha^{\overline{\rm{MS}}}_s(M_Z)=0.13\pm 0.005\pm0.03$ by the CLEO Collaboration~\cite{cleo} and $\alpha^{\overline{\rm{MS}}}_s(M_Z) =0.1224\pm 0.0039$ from a jet shape analysis~\cite{jet}.

\subsubsection{perturbative properties of $R_3$}

The above results indicate that the four-loop prediction for $R_3$ under various scale settings are close to each other. However, we shall show that the perturbative series for $R_n$ behaves quite differently using various scale settings. The convergence of the series is the key criterion for the reliability for a pQCD prediction -- determining which scale setting is the best for obtaining the most accurate prediction at a given fixed order. Moreover, a fast pQCD convergence means we need less loop calculations to achieve the same precision goal.

\begin{table}[htb]
\centering
\begin{tabular}{ c c c c c c c c}
\hline
 & ~~LO~~ & ~~NLO~~ & ~~N$^2$LO~~ & ~~N$^3$LO~~ & ~~$total$~~ \\ \hline
~~Conv.~~ & 0.04495 & 0.00285 & -0.00116 & -0.00033 & 0.04631 \\
PMC-I & 0.04290 & 0.00339 & -0.00002 & -0.00001 & 0.04626 \\
PMC-II & 0.04287 & 0.00350 & -0.00004 & -0.00002 & 0.04631 \\
PMS  & 0.04603 & 0.00010 & 0.00013 & 0.00008 & 0.04634 \\
\hline
\end{tabular}
\caption{The contributions of each loop-terms (LO, NLO, N$^2$LO and N$^3$LO) to the total four-loop prediction for $R_3$, in which the conventional scale setting (Conv.), the PMC-I, PMC-II and the PMS are adopted for setting the scale. $Q=31.6$ GeV and $\mu_0=Q$. } \label{Rorder}
\end{table}

To illustrate the pQCD convergence, we present the contributions of each loop-terms to the total four-loop estimate $R_3$ in Table \ref{Rorder}, in which the conventional scale setting, the PMC-I, the PMC-II and the PMS are adopted for setting the scale, respectively. Table \ref{Rorder} shows that the best pQCD convergence is achieved by PMC, in contrast to the moderate pQCD convergence of the conventional scale setting. The convergence of PMS oscillates; i.e,. its LO estimate is similar to that of conventional scale setting or PMC, but the results at NLO, N$^2$LO and N$^3$LO fail to show convergent behavior; i.e., $R^{\rm LO}_{3,{\rm PMS}} \gg R^{\rm NLO}_{3,{\rm PMS}} \sim R^{\rm N^2LO}_{3,{\rm PMS}} \sim R^{\rm N^3LO}_{3,{\rm PMS}}$ with $R^{\rm N^{2}LO}_{3,{\rm PMS}} > R^{\rm NLO}_{3,{\rm PMS}}$. This behavior is understandable, for the conventional scale setting, the pQCD convergence is guaranteed directly by the $\alpha_s$ suppression; for PMC, it is due to the elimination of divergent renormalon terms in addition to the $\alpha_s$ suppression; while, for PMS, its pQCD convergence should be an accidental, since the PMS scale is determined by requiring the estimate to be steady over the changes of renormalization scheme and scale, i.e. the local RGI.

It is helpful to be able to estimate the ``unknown" higher order pQCD corrections. The conventional error estimate obtained by varying the scale over a certain range is not reliable, since it only partly estimates the non-conformal contribution but not the conformal one. In contrast, after PMC and PMS scale setting, the scales are optimized and cannot be varied; otherwise, one will explicitly break the (standard/local) RGI which leads to an unreliable prediction. Thus, we will adopt another more conservative practice for the error analysis; i.e. to take the uncertainty to be the last known perturbative order. More explicitly, the perturbative uncertainty at the $(n+1)$-order is $\left(\pm |{\cal C}_{n} a^{n+1}_s|_{\rm MAX} \right)$, where both ${\cal C}_{n}$ and $a_s$ are calculated by varying the initial scale to be within the region of $[Q/2,2Q]$ and the symbol ``MAX'' stands for the maximum value of $|{\cal C}_{n} a^{n+1}_s|$ within this region. This treatment is natural for PMC, since after PMC scale setting, the pQCD convergence is ensured and the only uncertainty is from the last term due to the unfixed PMC scale at this particular order.

\begin{figure}[tb]
\centering
\includegraphics[width=0.6 \textwidth]{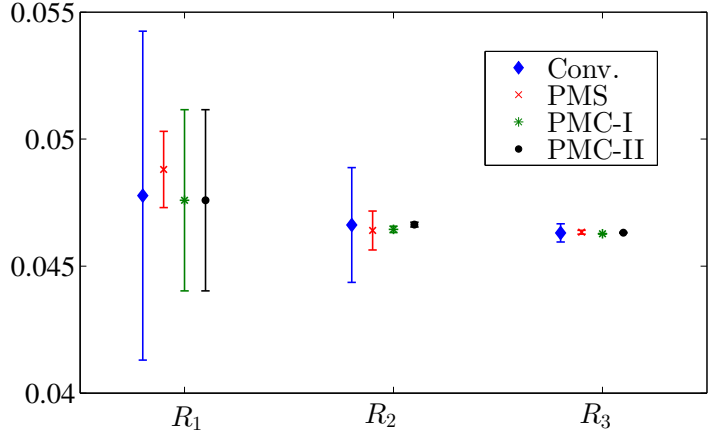}
\caption{Results for $R_n$ ($n=1,2,3$) together with their errors $\left(\pm |{\cal C}_{n}a^{n+1}_s|_{\rm MAX} \right)$ at $Q=31.6 $ GeV. The diamonds, the crosses, the stars and the big dots are for conventional scale setting (Conv.), PMS, PMC-I and PMC-II, respectively. } \label{error}
\end{figure}

The errors for the conventional and the PMC scale settings are displayed in FIG. \ref{error}. The predicted error bars from ``unknown" higher-order corrections quickly approach their steady points for PMC and PMS scale settings. The error bars provide a consistent estimate of the ``unknown" QCD corrections under various scale settings; i.e., the exact value for the ``unknown" $R_{n}$ ($n=2$ and $3$) are well within the error bars predicted from the one-order lower $R_{n-1}$. There is only one exception for PMS, whose $R_{2,3}$ is well outside the region predicted from $R_1$.

\subsection{$\Gamma(H \to b\bar{b})$ up to four-loop level}
\label{sec:htobb}

The decay width of $H\to b\bar{b}$ reads
\begin{eqnarray}
\Gamma(H\to b\bar{b})=\frac{3G_{F}M_{H}m_{b}^{2}(M_{H})} {4\sqrt{2}\pi}(1+\tilde{R}_n),
\end{eqnarray}
where $G_{F}$ is the Fermi constant, $M_{H}$ is the Higgs mass and $m_{b}(M_{H})$ is the $b$-quark $\overline{\rm MS}$ running mass. Up to $(n+1)$-loop correction, $\tilde{R}_n=\sum_{i=0}^{n}\tilde{\cal C}_i a^{i+1}_s(M_{H})$. At present, it has been calculated up to four-loop level, i.e. for $\mu_0=M_H$, we have~\cite{cor9}
\begin{eqnarray}
\tilde{R}_3 &=&5.6667 a_s(M_{H})+ (35.94-1.359n_f) \; a_s^2(M_{H}) \nonumber\\
&& + (164.14-25.77 n_f+0.259n_f^2) \; a_s^3(M_{H}) \nonumber\\
&& +(39.34 -220.9n_f+9.685 n_f^2-0.0205 n_f^3)\; a_s^4(M_{H}). \nonumber
\end{eqnarray}

\begin{table}[htb]
\centering
\begin{tabular}{cccccccccc}
\hline
 & ~~$\tilde{R}_1$~~ & ~~$\tilde{R}_2$~~ & ~~$\tilde{R}_3$~~ & ~~$\tilde{\kappa}_1$~~ & ~~$\tilde{\kappa}_2$~~ & ~~$\tilde{\kappa}_3$~~ \\ \hline
~~Conv.~~ & 0.24117 & 0.24314 & 0.24175 & $18.20\%$ & $0.82\%$ & $-0.57\%$ \\
PMC-I & 0.24890 & 0.24099 & 0.24119 & $21.99\%$ & $-3.18\%$ & $0.08\%$ \\
PMC-II & 0.24890 & 0.24104 & 0.24094 & $21.99\%$ & $-3.16\%$ & $-0.04\%$ \\
PMS  & 0.25581 & 0.24068 & 0.24125 & $25.38\%$ & $-5.91\%$ & $0.24\%$ \\
\hline
\end{tabular}
\caption{Numerical results for $\tilde{R}_n$ and $\tilde{\kappa}_n$ with various QCD loop corrections under the conventional scale setting (Conv.), PMC-I, PMC-II and PMS, respectively. The value of $\tilde{R}_0=0.20403$ is the same for all scale settings. $\mu_0=m_H$. }\label{tR}
\end{table}

\begin{table}[htb]
\centering
\begin{tabular}{cccccc}
\hline
 & ~~LO~~ & ~~NLO~~ & ~~N$^2$LO~~ & ~~N$^3$LO~~ & ~~$total$~~ \\ \hline
~~Conv.~~ & 0.20358 & 0.03761 & 0.00194 & -0.00138 & 0.24175 \\
PMC-I & 0.22658 & 0.02486 & -0.00908 & -0.00117 & 0.24119 \\
PMC-II & 0.22658 & 0.02500 & -0.00942 & -0.00123 & 0.24093 \\
PMS  & 0.23949 & 0.00061 & 0.00160 & -0.00046 & 0.24124 \\
\hline
\end{tabular}
\caption{The contributions of each loop-terms (LO, NLO, N$^2$LO and N$^3$LO) to the total four-loop prediction for $\tilde{R}_3$, in which the conventional scale setting (Conv.), the PMC-I, PMC-II and the PMS are adopted for setting the scale. $\mu_0=M_H$. } \label{Higgsbborder}
\end{table}

Following standard procedures, we can determine the results of $\tilde{R}_n$ and $\tilde{\kappa}_n$ (its definition is similar to $\kappa_n$ defined in Eq.(\ref{kappa0})) up to four-loop level under various scale settings, which are presented in Table.\ref{tR}. The contributions of each loop-terms (LO, NLO, N$^2$LO and N$^3$LO) to the total four-loop prediction for $\tilde{R}_3$ are presented in Table \ref{Higgsbborder}. At the four-loop level, the prediction for $H\to b\bar{b}$ under various scale setting are consistent with each other due to better pQCD convergence for all the scale settings. We also found the pQCD convergence of PMS is questionable, and its prediction of $\tilde{R}_2$ is also outside the region prediction from $\tilde{R}_1$. As an application, we obtain
\begin{eqnarray}
\Gamma(H\rightarrow b\bar{b})&=& 2389.48\; {\rm KeV}, \quad (\rm{Conv.}), \\
\Gamma(H\rightarrow b\bar{b})&=& 2388.52\; {\rm KeV}, \quad (\rm{PMS}), \\
\Gamma(H\rightarrow b\bar{b})&=& 2388.41\; {\rm KeV}, \quad (\rm{PMC-I}), \\
\Gamma(H\rightarrow b\bar{b})&=& 2387.92\; {\rm KeV}, \quad (\rm{PMC-II}).
\end{eqnarray}

\begin{figure}[tb]
\centering
\includegraphics[width=0.6 \textwidth]{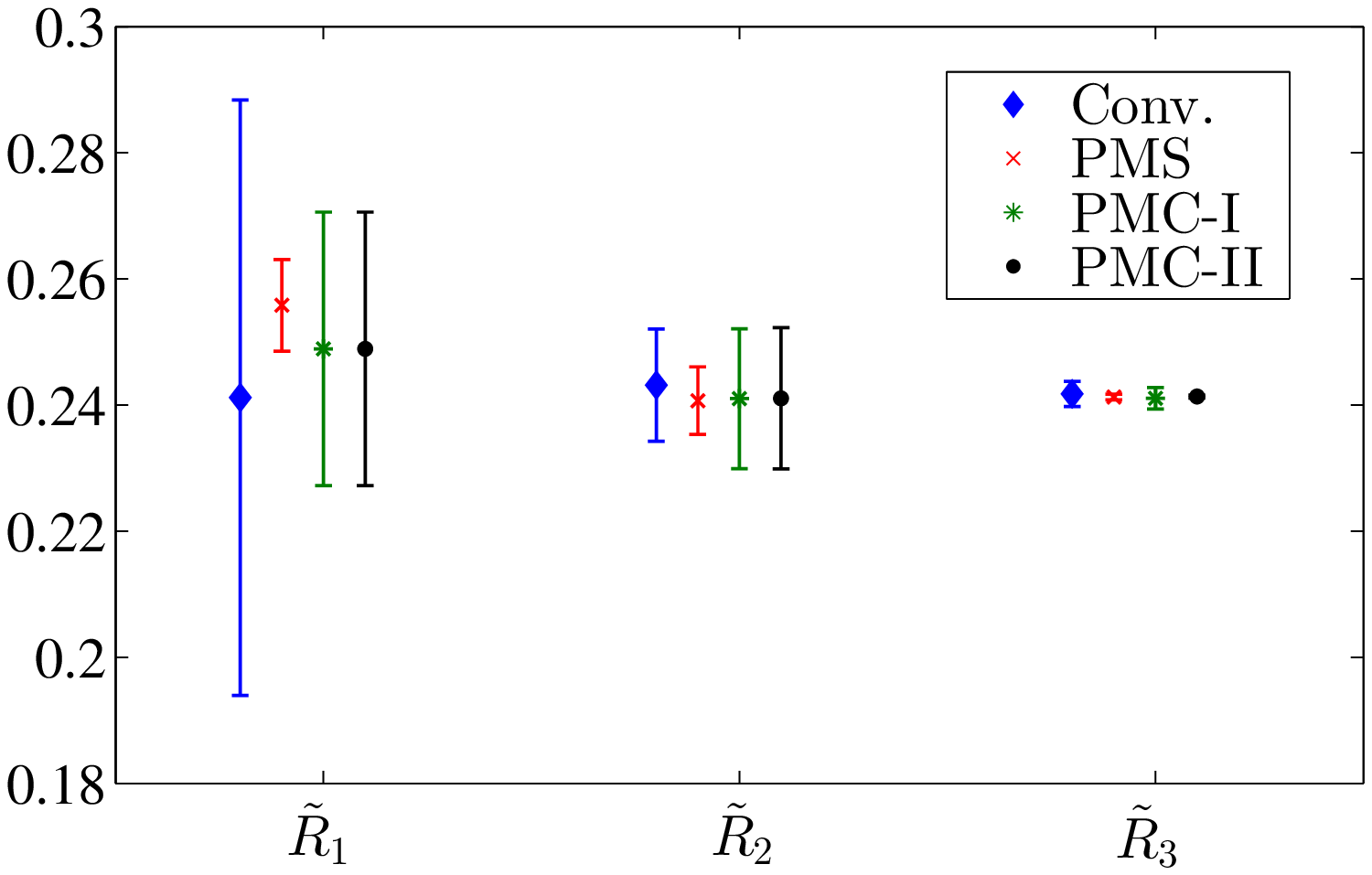}
\caption{Results for $\tilde{R}_n$ ($n=1,2,3$) together with their errors $\left(\pm |\tilde{\cal C}_{n}a^{n+1}_s|_{\rm MAX} \right)$ for $H\to b\bar{b}$. The diamonds, the crosses, the stars and the big dots are for conventional scale setting (Conv.), PMS, PMC-I and PMC-II, respectively. } \label{Higgserror}
\end{figure}

As in the case of $R_{e^+ e^-}$, we list the predicted errors $\left(\pm |\tilde{\cal C}_{n}a^{n+1}_s|_{\rm MAX} \right)$ for $\Gamma(H\to b\bar{b})$ for conventional scale setting, the PMC and the PMS in FIG. \ref{Higgserror}, where both $\tilde{\cal C}_{n}$ and $a_s$ are calculated by varying $\mu_0 \in [m_H/2,2m_H]$ and the symbol ``MAX'' stands for the maximum value of $|\tilde{\cal C}_{n} a^{n+1}_s|$ within this region. In the case of PMS, the values for $\tilde{R}_{2,3}$ are outside the predicted error bar from $R_1$. In the present case, conventional scale setting also performs well at fourth order as indicated by Table \ref{Higgsbborder}.

\section{Summary}
\label{sec:summary}

It is conventional to assume the renormalization scale in pQCD calculations to be equal to a typical momentum transfer of the process and varies it over an arbitrary range. This leads to an arbitrary systematic error for the fixed-order pQCD predictions. Moreover, the conventional method based on a guessed scale can lead to incorrect predictions when it is applied to QED processes. In principle the error can be suppressed by including more-and-more QCD loop corrections. However, this cannot be done in practice since the perturbative series inevitably diverges as $n! \beta^n \alpha_s^n$ at high orders due to renormalon terms.

It is clearly important to set the renormalization scale in a fundamental way consistent with the principles of renormalization group. The most critical criterion is that a prediction for a physical observable cannot depend on a theoretical convention such as the choice of renormalization scheme or the (initial) scale. This RGI principle is satisfied by the usual Gell Mann-Low scale setting used for precision QED predictions -- the QED scale is unambiguous, and the resulting high precision QED predictions are the same in any scheme at any finite order.

As we have shown in this review, the same RGI principle is satisfied for non-Abelian gauge theory when one uses PMC scale-setting. All terms in the pQCD series involving the $\beta$ function are absorbed into the running coupling order-by-order. The size of the PMC scale at each order also determines the effective number of contributing flavors $n_f$, just as in QED. The resulting coefficients of the pQCD series at any order using the PMC method are thus identical to that of the corresponding conformal theory with $\beta=0$ and are thus scheme independent. Unlike conventional scale setting, the divergent renormalon terms are eliminated.

The PMC thus provides a way to determine the optimal scale of the coupling constant for any QCD process via a systematic, scheme-independent and process-independent way. The PMC can also be applied to problems with multiple physical scales. For example, the subprocess $q \bar q \to Q\bar Q$ near the quark threshold involves not only the subprocess scale $\hat s \sim 4 M^2_Q$ but also the scale $v^2 \hat s$ which enters the Sudakov final-state corrections~\cite{sjbQQ}, where $v$ is the $Q \bar Q$ relative velocity. In the case of the top quark forward-backward asymmetry via the channel $p \bar p \to t \bar t X$, the application of the PMC reduces the difference between Tevatron measurements and the NLO pQCD predictions from $3$ standard deviations to about $1\sigma$~\cite{pmc10}, which agrees well with very recent measurement done by D0 collaboration~\cite{exptt}. In the recent paper which applies the PMC to the top-quark charge asymmetry at the LHC up to N$^2$LO level, it has been found that the PMC predictions are also in good agreement with the available ATLAS and CMS data~\cite{Wang:2014sua}. The critical feature of the PMC is that the renormalization scale that appears in the diagrams that interfere and produce the $t \bar t$ asymmetry are enhanced in QCD since those amplitudes have a smaller renormalization scale than the Born term. The same pattern of renormalization scales is also apparent in the $\mu^+ \mu^-$ asymmetry in the QED process $e^+ e^- \to \mu^+ \mu-$.

We have also discussed an alternative procedure, the PMS, which implements a local version of RGI, and we have given a detailed comparison of PMC and PMS predictions for two quantities $R_{e+e-}$ and $\Gamma(H\to b\bar{b})$ up to four-loop order in pQCD. At the four-loop level, the PMC and PMS predictions for $R_{e+e-}$ and $\Gamma(H\to b\bar{b})$ agree with conventional scale setting, and each of them show quite small scale dependences. However, the PMC prediction shows the fastest convergence to its four-loop value. The convergence of the PMS and PMC behave quite differently: as shown in Tables \ref{Rorder} and \ref{Higgsbborder}, the pQCD convergence is questionable for PMS. Worse, PMS scale setting disagrees with Gell Mann-Low scale setting when applied to QED and gives unphysical results for jet production in $e^+ e^-$ annihilation.

The PMC satisfies all self-consistency conditions deduced from RGI. The PMC also underlies CSRs between observables~\cite{Brodsky:1994eh}, such as the generalized Crewther relation~\cite{crew}. The PMC predictions have optimal pQCD convergence and are scheme and scale independent at any fixed order; any residual dependence on the choice of initial scale is highly suppressed, even for lower-order corrections. Thus in this sense, the PMC satisfies one of the requirement, i.e. Eq.(\ref{pmsstart1}), of PMS. The value for the effective number of flavors $n_f$ is set according the magnitude of the PMC scale just as in QED, thus eliminating another traditional ambiguity of pQCD. The PMC fixes the renormalization scale correctly at each order of pQCD via the $\beta$-terms than govern its running behavior via RGE, while the PMS only predicts an overall effective scale for the whole process.

We have suggested two approaches, PMC-I and PMC-II, to achieve the goal of PMC. It has been demonstrated that these two all-orders PMC approaches are equivalent to each other at the level of conformality and are equally viable PMC procedures~\cite{Bi:2015wea}. The PMC-I implementation is a direct extension of the BLM approach, whereas the PMC-II provides additional theoretical improvements; in addition, it can be readily automatized using the $R_\delta$-scheme. By construction, both the PMC-I and PMC-II satisfy all of the principles of the renormalization group, thus providing scale-fixed and scheme-independent predictions at any fixed order. Those two implementations of PMC differ, however, at the non-conformal level, by predicting slightly different RG scales of the running coupling. This difference arises due to different ways of resumming the non-conformal terms, but this difference decreases rapidly when additional loop corrections are included.

The key step of PMC-II is to use the pattern generated by the RG-equation and its degeneracy relations to identify which terms in the pQCD series are associated with the QCD $\beta$-function and which terms remain in the $\beta=0$ conformal limit. The $\beta$-terms are then systematically absorbed by shifting the scale of the running coupling at each order, thus providing the PMC scheme-independent prediction. The recursive patterns and degeneracy relations between the $\beta$-terms at each order are essential for carrying out this procedure. The implementation of PMC-II illuminates how the renormalization scheme and initial scale dependence are eliminated at each order. These advantages shows the PMC-II is theoretically robust and is the preferred method for practical implementations of PMC.

In addition to the examples discussed here, other PMC applications can be found in Refs.~\cite{pmcapp3,pmcapp4,pmcapp5,pmcapp6,pmcapp7,pmcapp8,Shen:2015cta}. The predicted error bars for ``unknown" higher-order corrections under the PMC scale setting quickly approaches a steady point. Thus one obtains the most accurate and optimal fixed-order estimate at any known order. An analogous method could be used for quark mass renormalization in pQCD: all terms associated with mass renormalization should be summed into the running mass order-by-order.

The {\it ad hoc} systematic error usually assigned to pQCD predictions is thus unnecessary and can be eliminated. The PMC, with its solid physical and theoretical background, greatly improves the precision of SM tests, and it can be applied to a wide variety of perturbatively-calculable collider and other processes.

\begin{acknowledgments}

Part of the development of the PMC was done in collaboration with Leonardo Di Giustino and Xu-Chang Zheng. We thank them, and also Andrei Kataev and Paul M. Stevenson for helpful discussions.

This work was supported in part by the Program for New Century Excellent Talents in University under Grant No.NCET-10-0882, National Natural Science Foundation of China under Grant No.11275280, the Fundamental Research Funds for the Central Universities under Grant No.CDJZR305513, the Department of Energy Contract No.DE-AC02-76SF00515 and by the Danish National Research Foundation under Grant No.DNRF90. SLAC-PUB-15953, CP3-Origins-2014-014 and DIAS-2014-14.

\end{acknowledgments}

\end{document}